\documentclass[aps,prd,nofootinbib,11pt]{revtex4} 

\usepackage[utf8]{inputenc} 
\usepackage{lmodern}  
\usepackage{amsmath} 
\usepackage{amsfonts}  
\usepackage{amssymb}  
\usepackage{graphicx} 
\usepackage{todonotes} 
\usepackage[english]{babel} 
\usepackage[caption=false]{subfig} 
\usepackage[pdfauthor="Lisa Glaser and Sebastian Steinhaus",pdftitle="Quantum Gravity on the Computer-Impressions of a workshop"]{hyperref}

\begin{document}
\title{Quantum Gravity on the computer: Impressions of a workshop}
\date{\today}
\author{Lisa Glaser and Sebastian Steinhaus}
\affiliation{Radboud University, Nijmegen, Netherlands and Perimeter Institute, Waterloo, Canada}

\begin{abstract}

Computer simulations allow us to explore non-perturbative phenomena in physics.
This has the potential to help us understand quantum gravity.
Finding a theory of quantum gravity is a hard problem, but in the last decades many promising and intriguing approaches that utilize or might benefit from using numerical methods were developed.
These approaches are based on very different ideas and assumptions, yet they face the common challenge to derive predictions and compare them to data.
In March 2018 we held a workshop at the Nordic Institute for Theoretical Physics (NORDITA) in Stockholm gathering experts in many different approaches to quantum gravity for a workshop on ``Quantum gravity on the computer''.
In this article we try to encapsulate some of the discussions held and talks given during this workshop and combine them with our own thoughts on why and how numerical approaches will play an important role in pushing quantum gravity forward.
The last section of the article is a road map providing an outlook of the field and some intentions and goalposts that were debated in the closing session of the workshop.
We hope that it will help to build a strong numerical community reaching beyond single approaches to combine our efforts in the search for quantum gravity.
\end{abstract}
\maketitle

Quantum Gravity is one of the big open questions in theoretical physics.
Despite recent successes in particle physics and cosmology, most notably the discovery of the Higgs boson and the direct detection of gravitational waves, we are still lacking a consistent description of physics from smallest to largest scales that reconciles gravity and the quantum nature of matter.
Possible signatures and effects of quantum gravity are numerous, from singularities in the early universe and black holes to the size and origin of the cosmological constant.
In addition to these fundamental issues, one might hope that future experiments could reveal other traces of quantum gravity.
Hence it is of utmost importance to push the development of quantum gravity approaches to a point where they make reliable predictions, which will allow us to verify or falsify theories.
 
In the last decades many promising non-perturbative approaches to describe space-time at the smallest scales have been developed,  (causal) dynamical triangulations~\cite{Ambjorn_Loll_1998,ambjorn_nonperturbative_2012}, causal set theory~\cite{Bombelli_Lee_Meyer_Sorkin_1987,Sorkin_2003}, group field theory~\cite{Freidel:2005qe,Oriti:2007qd} / tensor models~\cite{Rivasseau:2011hm,Rivasseau:2016zco,Gurau:2016cjo}, loop quantum gravity~\cite{cbook,thomasbook}, noncommutative geometry~\cite{Connes_1994}, spin foam models~\cite{Perez:2012wv,cfbook}, and others.
All of these postulate discrete structures that serve as a truncation on the number of degrees of freedom and allow for well-defined non-perturbative dynamics, akin to lattice gauge theories.
Previous research, in which these models are substantially simplified to be computable, has lead to impressive results, e.g. the resolution of the Big Bang singularity in loop quantum cosmology as a Big Bounce~\cite{PhysRevLett.96.141301}.
However, in order to make predictions for the full theory beyond simplifications and symmetry reduced models, we have to explore  their deep non-perturbative regime.
The bottleneck in this is the development of numerical techniques that allow us to efficiently extract results from the models, e.g. expectation values of observables and characteristics of different phases of the theory.
Encouraging developments have been made in recent years and the purpose of our workshop was to compare these across different quantum gravity approaches.

Within the last 30 years computers have revolutionized our lives and the way science is done.
While the very first physics computer simulations were 2d Ising models with $8 \times 8$ sites, the technology and its applications have evolved rapidly: today's high performance simulations can predict the gravitational waves emitted by two colliding black holes or neutron stars~\cite{pretorius_evolution_2005} and explain the masses of hadrons using lattice QCD~\cite{durr_ab_2008}.
These developments have slowly percolated into the quantum gravity community, and have given rise to mainly computational approaches to the problem, such as (causal) dynamical triangulations.
In these approaches the path integral for dimensions larger than two is too complicated to be tackled analytically, but numerical methods, adapted from QCD and statistical mechanics, show how a ground state with macroscopic features emerges~\cite{ambjorn_nonperturbative_2012}.

Other approaches have followed this example: in causal set theory Monte Carlo simulations are used to explore the space of all possible partial orders~\cite{henson_onset_2015}, which includes all geometries but also highly non-manifold like structures, and more recently to compare the prediction of a fluctuating cosmological constant to cosmological data~\cite{zwane_cosmological_2017}.
And in spin foam models numerical methods are indispensable to study the dynamics of spin foams with many degrees of freedom, e.g.~via the means of coarse graining / renormalization~\cite{Dittrich:2014ala,Bahr:2016hwc}.
Moreover, calculating the fundamental spin foam amplitudes also requires numerical techniques~\cite{Dona:2017dvf}.

In the workshop we brought together experts on these approaches to discuss recent developments in quantum gravity on the computer.
During the discussion two broad clusters of topics emerged; observables that we can measure and how we can reliably measure them, and numerical methods that are efficient for the different approaches.

In this article we would like to summarize these discussions and distill their main ideas.
We hope this will serve as a record of this workshop and a reference point for the current development of the field.

In the first section of this article we begin with a brief introduction to the various approaches discussed during the workshop.
The rest of the section is split into three subsections, where we discuss subtleties in defining the theories on the computer in subsection \ref{subsec:understand}, interesting observables in subsection \ref{subsec:obs} and numerical methods in quantum gravity in subsection \ref{subsec:num}.
In 
section \ref{sec:roadmap} we summarize the road map discussion of the last day and try to map goalposts and aspirations for the community.
A list of participants, slides and posters can be found on the website \href{nordita.org/qg2018}{nordita.org/qg2018}.

\section{Approaches, observables and numerical methods}

\subsection{Introduction to various approaches to quantum gravity}

Throughout this article we use different theories to exemplify the issues we want to discuss, as a reminder let us give a quick overview over frequently mentioned theories and their salient aspects.
In (causal) dynamical triangulations the path integral over geometries is regularized by introducing a triangulation, this has been explored analytically for two dimensional geometries and through simulations in two, three and four dimensions~\cite{ambjorn_nonperturbative_2012}.
The sum over geometries is implemented by summing over all possible triangulations, where the size of the simplices is kept fixed.
In dynamical triangulations, sometimes also called euclidean dynamical triangulations to distinguish it from causal dynamical triangulations (CDT), these simplices are equilateral, with all edges having the same length.
In causal dynamical triangulations the time like edges of the simplices have a different edge length, and a time-foliation of the geometries is enforced.
This leads to a very different ensemble of geometries in the path integral, and in particular suppresses changes in the topology which lead to degenerate behavior in euclidean triangulations.
In both approaches the simulations use a simplicial version of the Regge action~\cite{Regge_1961} to weight the  geometries.

In two dimensions dynamical triangulations can be solved using so called matrix models.
They give probability distributions for $N \times N$ random variables -- thus also called random matrices --, where the matrices are invariant under the conjugation of the unitary group. The action of these models then consists of matrix invariants, e.g.~the trace of a product of three matrices. This theory can be expanded in a sum over Feynman (ribbon) diagrams, where each diagram is dual to a discrete two-dimensional surface, e.g.~a triangulation if the interaction term is three-valent~\cite{DiFrancesco:1993cyw}.
Tensor models were developed to explore this method in higher dimensions.
Instead of integrating over random matrices and thus obtaining two dimensional surfaces, here the integrals are over higher order random tensors  with an action consisting of tensor invariants, thus creating surfaces in higher dimensions~\cite{Gurau:2016cjo,Rivasseau:2016zco}.

In several ways, group field theory (GFT)~\cite{Freidel:2005qe,Oriti:2007qd} is similar to tensor models. Using the same order interaction vertices, the combinatorics of the Feynman graphs of group field theories and tensor models agree. However, in addition to the combinatorics, the Feynman diagrams carry group theoretic data encoding a discrete geometry. The fields of the theory are defined on several copies of the underlying symmetry group. Crucially this group manifold is not related to a space-time manifold. Instead, space-time is supposed to emerge from field excitations, e.g.~as a condensate~\cite{Gielen_Oriti_Sindoni_2013}.
Group field theories are closely related to loop quantum gravity and spin foam models, e.g.~group field theories can be constructed whose Feynman diagrams are given by spin foam amplitudes~\cite{Oriti:2014uga}. As for quantum field theories, the consistency of GFTs is investigated through renormalization~\cite{Carrozza:2016vsq}.

Spin foam models~\cite{Perez:2012wv,cfbook} are a path integral approach to quantum gravity sometimes also referred to as covariant loop quantum gravity. Similar to previously described approaches, spin foams regularize the gravitational path integral by introducing a discretisation, a 2-complex, which is frequently chosen to be dual to a triangulation. The discrete geometry is again encoded in group theoretic data. For a given 2-complex, the path integral is implemented by summing over this data weighted by spin foam amplitudes. A priori, there is no rule determining which 2-complex to choose for a particular calculation, and generically the results depend on this. One way to address this is by also summing over all possible 2-complexes~\cite{Rovelli:2010qx}, which is systematically implemented in group field theory as discussed above. Alternatively, the refinement approach~\cite{Dittrich:2014ala,Bahr:2014qza}
aims at consistently defining the dynamics across various 2-complexes, e.g.~by relating the theories by identifying states on the boundaries of these complexes.

Among the theories discussed here, loop quantum gravity (LQG)~\cite{cbook,thomasbook} is the only approach aiming to canonically quantize gravity.
To this end space-time, which is assumed to be globally hyperbolic, is split into space and time.
Due to diffeomorphism symmetry, the theory is totally constrained, i.e.~the Hamiltonian itself is a sum of constraints, such that the dynamics amount to gauge transformations.
Moreover, these constraints form the so-called hyper surface deformation algebra.
The goal of LQG is to quantize this algebra of constraints via Dirac quantization.
To achieve this, one defines a kinematical Hilbert space, whose states do not satisfy the constraints, and constructs suitable constraint operators and an associated operator algebra.
Then, the final goal is to find the physical Hilbert space, i.e.~all states annihilated by the constraints. As an alternative method to tackle this issue, spin foam models have been developed as the ``covariant'' theory to LQG. While the two frameworks are closely related, e.g.~the boundary states of modern spin foam models are kinematical states of LQG, their connection is not completely understood~\cite{Alesci:2011ia,Thiemann:2013lka}.

In causal set theory (CST) space-time  is reduced to a partially ordered set.
The discrete events are related to each other only if they are causally connected~\cite{Bombelli_Lee_Meyer_Sorkin_1987}.
This leads to a minimal amount of structure assumed, which is why reconstructing space-time from a causal set is a complicated problem.
There are methods to recover manifold properties from a causal set, maybe the simplest is to recover time like distance between two events by counting the longest chain between them.
Recovering space like distances is more complicated, but still possible~\cite{Eichhorn:2018doy}, and we can even define a measure allowing us to identify local regions, in the sense of regions that are small compared to the curvature scale of the manifold~\cite{Glaser_Surya_2013}.
A causal set is considered to be manifold-like if it could have, with high likelihood, arisen from a statistical, so-called sprinkling, process on a given manifold (for a good definition and an algorithm to reconstruct the embedding see~\cite{Henson_2006}).

Modern string theory describes open and closed strings in $10+1$ dimensions~\cite{Zwiebach:2004tj}.
Since higher dimensions often lead to more trouble in computer simulations this has not extensively been explored numerically.
The old, bosonic, string theory, which describes the quantization of $2$d surfaces covered by strings, can be studied numerically~\cite{Polchinski:1998rq}.
In fact, this was one of the motivating examples for the dynamical triangulations approach.
This is often called non-critical string theory, and is an example of a theory that can be solved analytically but also explored using simulations~\cite{ambjorn_string_2016}.

One might debate whether noncommutative geometry really offers an approach to quantum gravity, or is purely a mathematical generalization of the concept of manifolds.
A compact Riemannian manifold can be expressed as an algebra of functions acting on a Hilbert space together with a Dirac operator, a so-called spectral triple.
Generalizing this description to allow for noncommutative function algebras then extended the space of geometries allowed~\cite{Connes_1994}.
While the original examples were concerned with infinite dimensional algebras it is also possible to construct finite matrix algebras that then converge towards continuum geometries in the limit of infinite matrix size.
These are the so called fuzzy spaces which have recently been proposed as possible states in the path integral for quantum gravity~\cite{Barrett_Glaser_2016,Glaser_2017}.

The asymptotic safety approach~\cite{Reuter:2012id} hinges on Weinberg's idea~\cite{Weinberg:1980gg} that quantum gravity, described as a quantum field theory, is non-perturbatively renormalizable, i.e.~possesses an interacting fixed point of the renormalization group flow in the ultraviolet described by a finite amount of relevant coupling constants. In practice, this hypothesis is investigated via the functional renormalization group~\cite{Wetterich:1992yh}, where one integrates out short scale degrees of freedom to derive an effective theory at larger scales. Generically, this operation cannot be performed in full generality and requires truncations, e.g.~only particular terms in the action,  called the theory space, are considered. To check whether signs of a fixed point persist once more interactions are allowed, the theory space is consistently enlarged.
Work in this theory is mostly done using analytic methods or computational algebra packages, thus not exactly qualifying it as a numerical approach. However it can play an important role in connecting continuum to discrete theories, and thus testing predictions.

\subsection{Subtleties in defining a theory (on the computer)} \label{subsec:understand}
In the past decades we have seen tremendous progress in the definition and development of non-perturbative approaches to quantum gravity.
While some of these approaches share similarities, e.g.~the use of discrete structures to calculate the non-perturbative regime, they are based on very different assumptions and key ideas about what a theory of quantum gravity should be.
This variety itself is an opportunity and should be embraced rather than antagonized, yet it arises due to one of the great weaknesses of quantum gravity, the lack of experimental data to guide development.
However a diverse set of approaches gives us the chance to uncover universal features across theories and  to reveal the consequences of their underlying assumptions.
To make the most of this chance it is indispensable to make an effort to better understand the theories and their connections to one another.

Since we rely on numerical simulations in order to compute results, e.g.~expectation values of observables, it would be ideal to know exactly how to choose the parameters of the theory, i.e.~coupling constants or size of the discretisaton, to reliably and efficiently get the  ``right'' answer.
A prime example is lattice QCD~\cite{Gupta:1997nd}, in which numerical methods provide accurate predictions, e.g.~for the hadron spectrum~\cite{durr_ab_2008}.
Two features are crucial for its success: its direct contact to experiments and the existence of a renormalizable continuum theory.
On the one hand{\color{red},} the renormalizability of the continuum theory, thanks to asymptotic freedom~\cite{Gross:1973id}, makes it possible to determine the dynamics, i.e.~the coupling constants, at different scales.
On the other hand, experimental data fixes the parameters of the theory and tells us, which scale is relevant for a particular process.
Naturally this does not imply that the simulations can be straightforwardly performed, but it allows practitioners of QCD to focus their efforts on specific regions in parameter space. In his talk, Jack Laiho described in detail the challenges one faces in lattice QCD calculations, in particular with respect to fermionic degrees of freedom.

Considering their importance for the success of lattice QCD it seems crucial to tackle the issues of renormalization, an effective continuum theory and contact to experiments in quantum gravity.
Here we understand renormalization in the Wilsonian sense~\cite{Wilson:1973jj}, as a scheme to relate theories defined at different scales.
Usually one orders the degrees of freedom according to scale, then integrates out those at shorter scales to derive an effective theory on larger scales, ultimately relating a microscopic dynamics to macroscopic physics.
Additionally, choice of parameters, ambiguities or the choice of discretisation in the microscopic, allegedly fundamental, theory might give rise to different continuum dynamics strongly affecting observable quantities.
We would summarize these as different phases of the theory.
Conversely by exploring this phase diagram we can identify regions of universal behaviour of the theory, unravel phase transitions and fixed points and hence check the consistency of the theory.

Finding a systematic framework that can relate theories at different scales in a background independent setting is a challenge.
In her talk, Bianca Dittrich described a thoroughly studied proposal in spin foam models based on the idea to relate theories by identifying the same physical transitions on different discretisations~\cite{Dittrich:2014mxa,Dittrich:2013xwa}, and thus scales, in order to find theories giving consistent answers. In particular she emphasized that consistency is indispensable for extracting predictions from the theory, e.g. expectation values of observables.
To make progress in this direction it is worthwhile to implement approximations and simplifications in order to cover a larger part of the parameter space with given resources.

\subsubsection{Relating to the continuum}
Closely related to the issue of renormalization is the question of a continuum limit or at least an effective continuum theory compatible with any particular discrete quantum gravity theory.
Ideally such a continuum theory should agree with general relativity in a suitable limit, but it might also reveal crucial deviations that experiments can search for.
One possible relation discussed at the workshop, was to compare the 3-volume correlations computed in CDT with an effective continuum theory. Interestingly this can also be studied in other approaches and explored using functional renormalization group techniques~\cite{Knorr_Saueressig_2018}.
However, special care is advised when comparing continuum theories and their discretisations, as relating numerical simulations to   analytic solutions can give rise to new subtleties.

A particularly interesting example is the bosonic string, as pointed out by Jan Ambj{\o}rn.
The bosonic string can be solved with analytic as well as numerical methods, however these two solutions do not necessarily agree.
The reason for this conundrum is an incompatibility of the renormalization procedures; the continuum theory used dimensional regularization, and hence did not generate certain terms that arose in the discrete theory. Repeating the continuum calculation using a different regularization scheme made it possible to match the continuum and discrete results~\cite{ambjorn_string_2016}.
This showcases how much care needs to be taken in mapping analytic and numerical results onto each other.
This illustrates that ``brute force'' applications of known methods may not be directly applicable in the context of quantum gravity.

\subsubsection{Approximations and simplifications}
Another particularly contentious issue is the use of approximations and simplifications in computer simulations.
The most obvious of these is that simulated models are necessarily much smaller than the real universe.
The space-time volume of our universe is about $10^{240}$ Planck volumes in size, which does not compare well to, e.g.~the size of $10^2$ Planck volumes currently examined in causal set theory.
Some theories do better but in general the size of the universe in current discrete approaches is of the order $10^0$ to $10^5$ discrete building blocks.
Of course simulating the entire universe from quantum gravity might be too ambitious, and it might suffice to simulate a small region of space-time that recovers general relativity semi classically.
The current best tests of general relativity limit corrections to appear on a scale below $47\mu$m~\cite{PhysRevD.98.030001}\footnote{This number is estimated by assuming that if extra dimensions of this size can not be experimentally excluded it gives a conservative upper limit on the scale at which quantum gravity would appear.}. 
Assuming we wanted to simulate a cube of space-time of this extend in all four dimensions we would need to simulate $\sim 10^{122}$ Planck volumes which is still out of reach by several orders of magnitude.
One might argue that it is only a question of time, and better code to improve this, but no matter how good our code will be, the size of our simulations will be limited by the need to build our computer within the universe, and out of atoms.
Hence careful reasoning and planning about how to best use our limited resources is an important part of pushing forward numerical quantum gravity.

Many current simulations, in particular those using Monte Carlo methods, use Wick rotated geometries and statistical physics methods that allow for faster convergence of the results. However, it is not clear how the theory is affected by these changes, e.g.~whether the ensemble with respect to which one samples geometries is significantly altered. Moreover, effects typical for quantum superpositions might be obscured by this choice. Conversely, in some approaches it is not clear how to define a Wick rotation in the first place.
The only way to control for these factors would be to find algorithms and implementations working with oscillating amplitudes. One such method are tensor network renormalization techniques~\cite{Levin:2006jai}, which on the other hand are limited by numerical cost, which increases with the complexity of the studied system.
A promising future direction might be simulating quantum systems on actual quantum computers. This could avoid the problem of complex phases and make it possible to explore superpositions of states.
Even disregarding these fundamental points, there are still other simplifications and limitations we need to include in our theories, and it is important to be aware of these and explore their limits.

More specifically, theory dependent examples of simplifications are the foliation in CDT, the restriction to particular geometric intertwiners in current spin foam simulations, and the $2$d orders in Causal set theory.
In CDT the simulations fix space-time to be foliated into constant time slices and to have a constant topology.
This limitation has proven necessary to suppress so called ``baby-universes'', which have been identified as the reason that dynamical triangulations are so irregular and do not show good continuum behavior in the simplest examinations.
However this limitation has been explored and challenged: a certain rescaling of the matrix model for 2d dynamical triangulation suppresses the baby universes and leads to the same behavior as CDT~\cite{ambjorn_relation_1999}. Also in more recent work, it was shown that simulations without a strict foliation, but still conserving a time-orientability condition, lead to a good continuum behavior in 2 and 3 dimensions~\cite{jordan_sitter_2013,jordan_causal_2013}.
These results are expected to also hold for 4d, however have not been tested yet there due to technical challenges.
Nevertheless, they lend some credibility to the claims that the foliation in CDT is a simplification that does not overly constrain the phase space of the model. Moreover, this foliation can be used to employ efficient algorithm, like the transfer matrix algorithm described in Andrzej G\"orlich's talk, see also section \ref{subsec:num}.
Additional hints for this come from recent results obtained in euclidean dynamical triangulations with an additional curvature term.
These simulations show a first order phase transition, but  it is conjectured that this transition ends at a critical point that could be in the same universality class as CDT~\cite{Laiho:2016nlp}.

Spin foams also come initially with a large theory space that is hard to explore in full generality.
Indeed calculating the fundamental amplitudes of the theory is analytically not possible and requires a lot of computational resources, even for a single building block~\cite{Dona:2017dvf,Dona:2018nev}.
Studying larger spin foams is systematically tackled in the framework of renormalization~\cite{Dittrich:2014ala} described in Bianca Dittrich's talk, where effective degrees of freedom at a coarser level are defined from the full amplitude without ad-hoc truncations. A suitable numerical scheme are so-called tensor network techniques~\cite{Levin:2006jai}, in which the system is rewritten as a contraction of a network of tensors, i.e.~multidimensional arrays.
The goal is to approximate said network by a coarser network efficiently by locally manipulating the tensors, e.g.~sorting degrees of freedom according to their relevance via a singular value decomposition. These methods are particularly useful for identifying different phases of the model, e.g.~in 2D analogue spin foam models~\cite{Dittrich:2013bza,Dittrich:2013voa,Dittrich:2014mxa,Dittrich:2016tys} and 3D lattice gauge theories~\cite{Dittrich:2014mxa,Delcamp:2016dqo},  where they revealed rich phase structures and phase transitions.
Benjamin Bahr presented a closely related, but less holistic ansatz suitable for studying 4D spin foams: the underlying idea is to restrict the theory space to specific geometric shapes, e.g.~cuboids~\cite{Bahr:2015gxa} or frusta~\cite{Bahr:2017eyi}, which are coarse grained by requiring agreement of expectation values of observables across discretisations.
Instead of a triangulation, the combinatorics of the foam are chosen to be hypercubic such that the coarse graining procedure can be staightforwardly iterated.
Integrating over all possible shapes for the polyhedra is computationally prohibitively expensive, the using the simpler cuboids allowed for calculating the first 4D RG flow of (restricted) spin foam models and revealed indications for a phase transitions and a UV-attractive fixed point~\cite{Bahr:2016hwc,Bahr:2017klw}. Similarly the spectral dimension in the cuboid case  showed signs of a phase transition, where one phase is characterized by a dimension of four~\cite{Steinhaus_Thurigen_2018}. 
Moreover, a candidate for a similar fixed point was also found in the frusta setting which extends the space of allowed geometries compared to the simpler cuboids~\cite{Bahr:2018gwf}.

As a last example, in most of the current explorations of the dynamics in CST the path integral is restricted to only sum over the so-called $2$d orders.
These are a subclass of causal sets that can always be embedded into a plane, and that are dominated by causal sets that could arise from sprinkling in $1+1$d Minkowski space.
Sumati Surya told us about these and their limitations, opportunities and possible extensions in some detail.
This has two practical reasons, one is that the class of $2$d orders is much smaller than that of all causal sets, and hence much easier to explore on the computer.
The class of all possible causal sets grows like $2^{N^2/4}$, and is dominated by the very non-manifoldlike Kleitman-Rothschild orders~\cite{Kleitman_Rothschild_1975}, numerically this dominance sets in for $N \gg 90$~\cite{henson_onset_2015}, which makes it very hard to explore in computer simulations.
The other reason is that the choice of $2$d orders immediately answers a number of questions one needs to debate before simulating causal sets, namely those concerned with how to pick the dimension of space-time, and hence the action to use in the simulations. Furthermore, it also allows us to store the causal set in a $2$d array and thus enables faster algorithms.
\begin{figure}
    \includegraphics[width=0.3\textwidth]{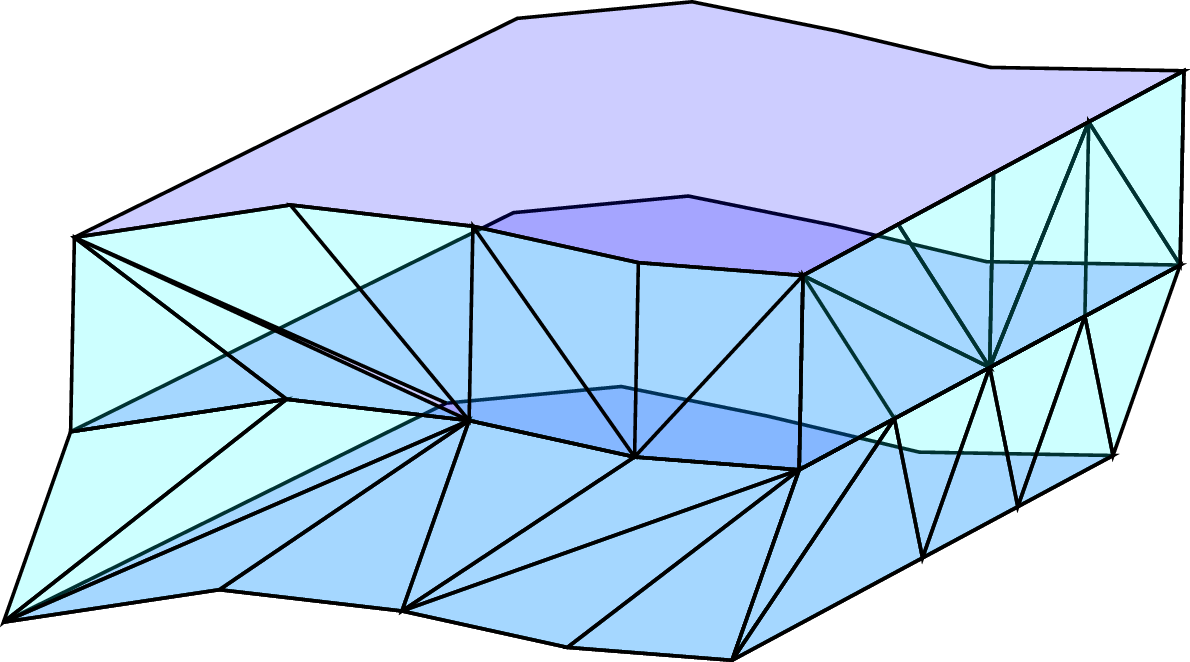}\hspace{30pt}
    \includegraphics[width=0.3\textwidth]{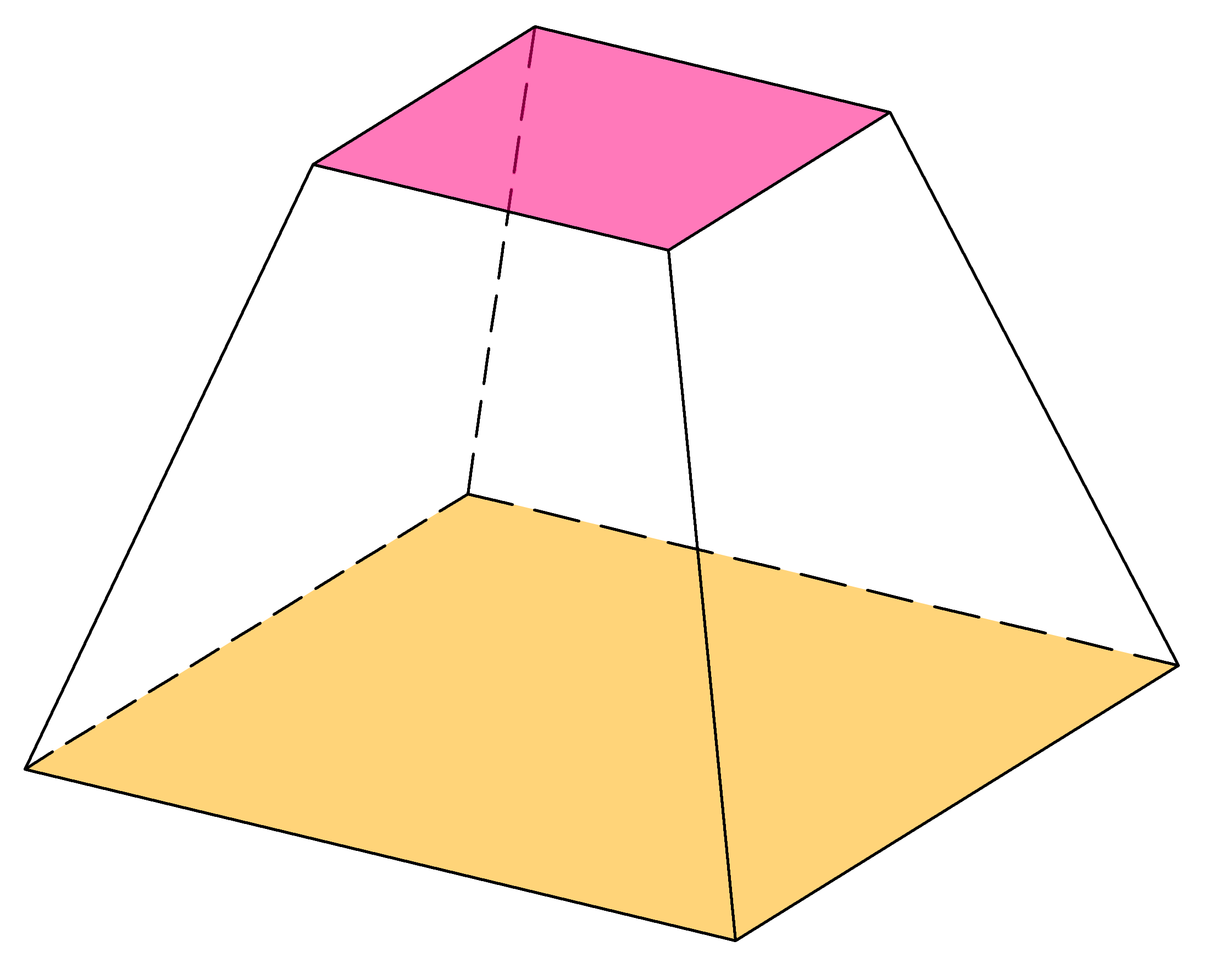}
    \includegraphics[width=0.3\textwidth]{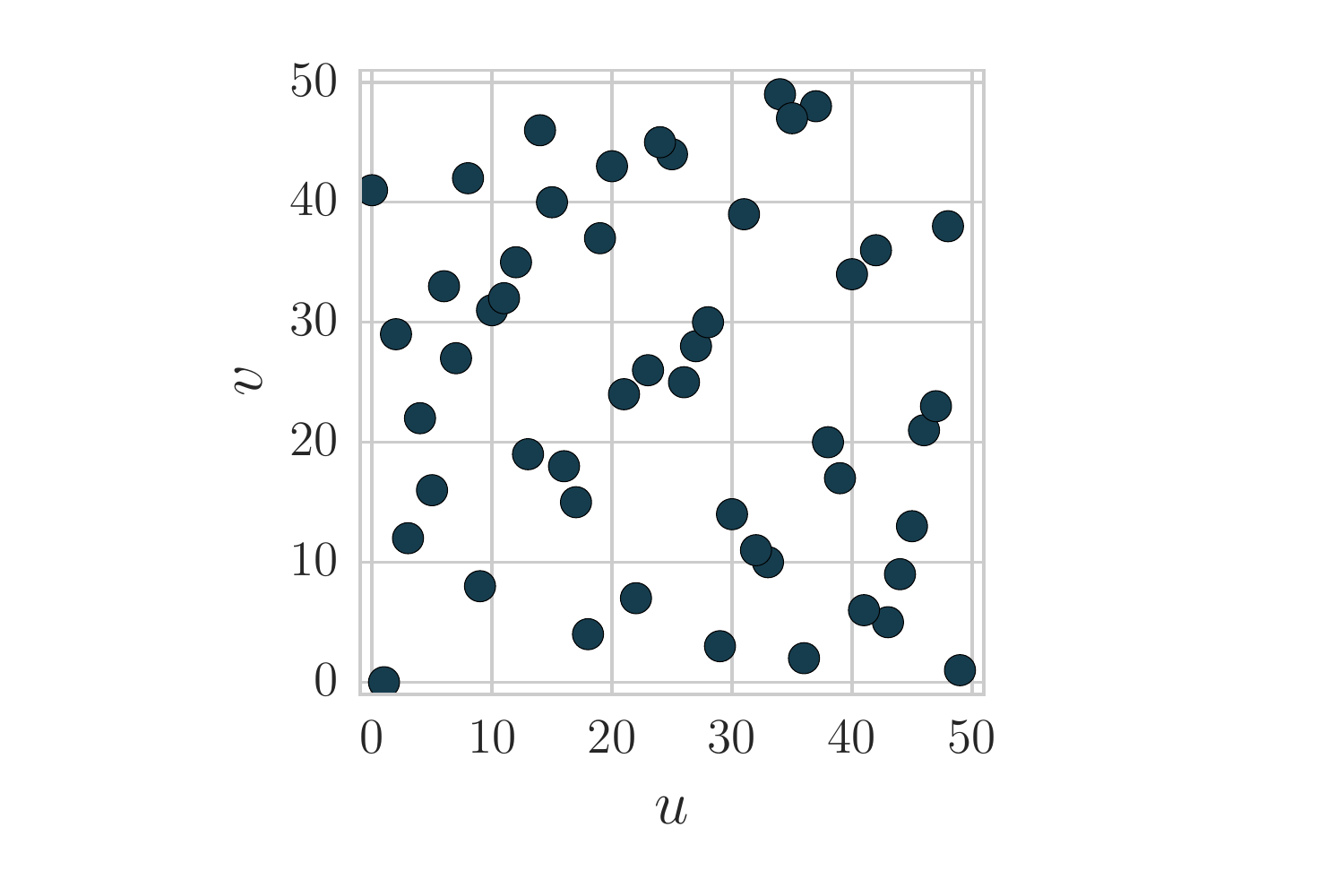}
    \caption{Examples of different simplifications used in the theories. From left to right we see a causal triangulation with a foliation, a square frustum for spin foams and a $2$d order causal set.}
    \label{fig:simplifications}
\end{figure}

The issue of limited numerical resources and necessary simplifications sheds a light onto the question how we can efficiently use them to reveal the properties of space-time and work towards making contact with experiments. Indeed the latter point is certainly difficult for a theorist.
Optimally we would like to study observables that are well-defined both in discrete and continuum theories, yet connecting these to observable physical effects is usually a harder question.
Thus, in order to deepen the connection between abstract quantum gravity theories and phenomenology, it is imperative that different quantum gravity approaches strive towards defining and studying the {\it same} observables. Then we can unveil similarities and differences between approaches that might stimulate development and realization of experiments capable of testing multiple theories at once.

Indeed there is great potential in studying observables in quantum gravity.
In the next section we present some proposals discussed during the workshop.

\subsection{Observables}\label{subsec:obs}

Many observables have been proposed to better understand quantum gravity.
One possible consequence of quantum gravity that can arise in different ways and was discussed at length in our workshop, is non-locality.
However, the meaning of non-locality depends heavily on the context it is discussed in, and it is not completely understood how these notions are related.
Even in classical general relativity locality is a subtle concept.
This is due to diffeomorphism invariance -- the fundamental symmetry of general relativity, which encodes the independence of physics under the choice of coordinate systems. As a result only diffeomorphism invariant quantities are physically relevant. For example, this condition severely complicates the definition of local subsystems in general relativity. Indeed, splitting systems into subsystems, e.g.~to compute the entanglement entropy between them, is highly topical, yet in gravity it must be defined in a diffeomorphism invariant way. Similar to the situation in (lattice) gauge theories, this can be achieved by introducing new degrees of freedom and symmetries on the boundary separating the subsystems~\cite{Donnelly:2016auv}.

\subsubsection{Non-locality in quantum gravity}
One facet of non-locality discussed at the workshop was in the context of effective quantum field theories.
The essential idea put forward by Knorr and Saueressig is to define an effective continuum theory for CDT~\cite{Knorr_Saueressig_2018}, where the terms and couplings in the effective field theory are chosen by comparing expectation values of the 3-volume covariance in both theories (for one specific value of parameters in CDT).
The theory they define contains non-local terms, in the sense that the associated operator is a product of the field (and its derivative) evaluated at different points in space-time\footnote{More precisely, the operator is the inverse d'Alembertian squared sandwiched by two Ricci scalars.}. It remains an open question whether similar relations hold once more observables are considered or when the parameters in CDT are changed. Nevertheless, the potential implications of such non-local terms are intriguing and it will be interesting to explore whether similar effective quantum field theories can be derived from other discrete quantum gravity approaches.
Non-locality can also arise in discrete theories.
Spin foam models and CDT can be regarded as (at least initially) local theories, since they assign amplitudes to each building block of the triangulation, where these amplitudes only depend on the variables attached to said building block. A priori only neighbouring building blocks are ``interacting'' via the variables they are sharing.
However, under coarse graining / renormalization generically non-local interactions will arise involving building blocks beyond nearest neighbours.

Conversely, in causal set theory non-locality is built into the theory from the beginning.
A causal set element is connected to all causally related elements, with nearest neighbours corresponding to elements close to the light-cone.
Since the light-cone in a generic space-time is non-compact, a causal set element, in an infinite causal set, would have infinitely many nearest neighbours.
Additional non-locality also arises through a regularization parameter in the d'Alembertian for a scalar field~\cite{Dowker_Glaser_2013, Glaser_2011, Aslanbeigi_Saravani_Sorkin_2014, Sorkin_2007}.
This parameter is introduced to dampen fluctuations in the discrete theory, in effect smearing the derivative operator over several layers.
This non-locality of a scalar field on a causal set, and of the causal set itself gives rise to phenomenological predictions which can be tested~\cite{ Belenchia_Benincasa_Liberati_2015, Belenchia_Benincasa_Martin-Martinez_Saravani_2016, Belenchia_Benincasa_Liberati_Marin_Marino_Ortolan_2016}.
On the other hand, we do not have any current observational evidence of non-locality, hence any non-local effects need to remain weak enough to not conflict with this.
For example when modelling the motion of a point particle through a causal set as traversing along the longest path, this introduces momentum diffusion above these stringent limits~\cite{Philpott_Dowker_Sorkin_2009}.

\subsubsection{Summing over topology}
Another point of contention between different theories is the question: If we sum over different geometries, should we hold their topology fixed, or should we sum over all possible topologies?
The first time this problem arose was in non-critical string theory, where the theory of strings requires a complete sum over all possible topologies of $2$d geometries.
If this is naively implemented it leads to a dominance by topologies with many handles and a divergent sum~\cite{Ambjorn_Durhuus_1997}.
However modern models, and a suitable renormalization, make it possible to calculate the sum over topologies.
In CDT the topology of space and time individually are fixed, the simulations restrict spatial topology to either be a sphere or a torus, and for numerical reasons time is treated as periodic in most simulations\footnote{With the exception of the results in~\cite{Cooperman_Miller_2014}, which broadly agree with the results found using periodic time.}.
Causal set theory on the other hand does not restrict the path integral in this way, it does not even require all partial orders in the path integral to be geometries.

In loop quantum gravity and other canonical formulations of quantum gravity the topology of space-time is usually fixed, since space-time is assumed to be globally hyperbolic in order to define the (3+1) split.
Indeed describing time evolution in a non-globally hyperbolic space-time is rather cumbersome.
In spin foams the issue is more subtle: any 2-complex that is compatible with given boundary data is in principle allowed.
This concerns non-trivial 4-dimensional topology but also includes the possibility to change spatial topology between initial and final state.
Whether one should sum over different topologies is debated in the literature~\cite{Rovelli:2010qx} and depends on the interpretation of the spin foam.
In the refinement approach~\cite{Dittrich:2014ala}, where the goal is to define a consistent theory across discretisations, one usually does not consider topology change.
Since the goal is to identify the same physical process across discretisations, it is natural to fix the topology of the boundary states\footnote{If the boundary consists of several parts, e.g.~initial and final state, their respective topologies can differ, but are kept fixed. Moreover it is not clear how to embed states of differing topologies into a common discretisation.}.
The topology in the bulk is usually kept fixed as well, mostly for convenience.
On the other hand it is frequently argued that one should sum over all possible spin foams, i.e.~all 2-complexes including all topologies.
The most suitable framework to consider this are group field theories~\cite{Oriti:2014uga}, in which the sum over spin foams appears as a perturbative expansion of Feynman diagrams generated by the action of the theory.
Whether this theory is well-defined depends on whether it is renormalizable as a quantum field theory~\cite{Carrozza:2016vsq}.
In the related tensor models the sum over topologies is well defined.
In particular they can identify which topologies dominate in the perturbative expansion in the so-called large-$N$ limit, where $N$ is the dimension of the tensor indices~\cite{Rivasseau:2011hm,Gurau:2016cjo}.

\subsubsection{Quantum cosmology}

One of the most promising routes for quantum gravity to make contact with experiments is cosmology. Quantum gravity effects may be revealed by future high precision experiments, e.g.~the dynamics of the early universe might have left imprints in the cosmic microwave background. Indeed, it is an exciting prospect to see how quantum gravity can reshape our understanding of the origin of the universe, and whether it can augment, replace, or derive the current paradigm of inflation~\cite{Linde:2007fr}, which successfully explains the (almost) homogeneity, isotropy and flatness of our universe.

However, contact to this cosmological sector is difficult for non-perturbative theories of quantum gravity.
While it is challenging for many approaches to define or model such a subsector of the theory, it is even more so to show how such a sector (plus fluctuations around it) could emerge dynamically.
This difficulty is exemplified by the difference between loop quantum cosmology (LQC) and cosmology in loop quantum gravity: in loop quantum cosmology, the system is symmetry reduced, e.g.~to a homogeneous and isotropic universe, at the classical level before quantization. In loop quantum gravity, this symmetry reduction is to be implemented at the quantum level and explored in different directions. 
The early symmetry reduction in loop quantum cosmology simplifies calculations considerably and allows for interesting tests.
For example LQC with an inflationary phase after the bounce predicts changes in the CMB power spectrum compared to other inflationary models~\cite{LQC_CMB}.
However there are strong arguments that the early reduction in symmetry might remove crucial information from the theory, hence to confirm the results of loop quantum cosmology it is vitally important to derive symmetry reduced models from the full theory.

Antonia Zipfel gave a nice overview of the current status of the relation between LQC and LQG: In loop quantum gravity this can be tackled directly by looking for suitably defined cosmological subsectors~\cite{Fleischhack:2015nda,Hanusch:2013jza}, e.g.~by translating homogeneity and isotropy conditions on the phase space of general relativity to loop quantum gravity~\cite{Beetle:2017qle}. While this procedure is mathematically robust and relates well to the full theory, it is hard to implement in a given model and only approximately recreates the symmetry.
A different idea is to study the evolution of coherent states, e.g.~peaked on homogeneous and isotropic space-times~\cite{Assanioussi:2018hee,Dapor:2017rwv}. Using these states one can derive an effective Hamiltonian, as the expectation value of the constraint with respect to these semi-classical states. However, it is a priori not clear  whether these coherent states are preserved under evolution. Another attempt to connect loop quantum gravity and loop quantum cosmology is called quantum reduced loop gravity~\cite{Alesci:2016gub}, which relies on the kinematical construction of full loop quantum gravity. Then a gauge fixing that restricts the spatial metric (and triads) to be diagonal is implemented. The symmetry reduction happens at the quantum level, where one only considers a dynamics which preserves the diagonal metric condition.
Yet another perspective on the difference between imposing symmetry reduction before or after quantization is given in the context of general relativity in radial gauge~\cite{Bodendorfer:2015aca}. In~\cite{Bodendorfer:2015qie} the two methods are closely compared, beginning at the level of the phase space in order to identify the variables in the reduced theory with suitable phase space functions in the full theory. This analysis is continued at the quantum level, where the subsectors of the theories and the properties of operators can be compared. While a qualitative match between both theories is achieved at the kinematical level, one finds quantitative and state-dependent differences in the scaling behaviour of operators and mismatches in their commutators. This suggests that the identification of subsectors needs to be improved further.

This problem, of how and where symmetry should be imposed arises in all non-perturbative approaches to quantum gravity and is dealt with in different ways.
Reducing the symmetry classically and then quantizing leads to interesting toy models, however it is important to test results obtained thus against results arising in the full non-perturbative regime.
In particular it would be fascinating if a non-perturbative path integral might give rise to a ground state that has some cosmological features.
This is the case in causal dynamical triangulations, where the ground state of simulations in one phase shares some characteristics with Euclidean de~Sitter space.
The average volume profile of the $3$-volumes, centered in time, measured in simulations assuming a spherical topology of space, matches the volume profile of euclidean de~Sitter~\cite{Ambjorn_Jurkiewicz_Loll_2004}. In addition to the 3-volume the authors also studied the covariance between 3-volumes at different time steps, which is highly peaked for the same time and drops off quickly for larger time steps.
The spectral dimension in this phase of the simulations also points at $4$ dimensional behavior~\cite{Ambjorn_Jurkiewicz_Loll_2005}.
This work has been extended to toroidal topology, where the volume profile becomes constant~\cite{Ambjorn:2017ogo}.
It can be argued that this creation of a de~Sitter volume profile is  a non-perturbative emergence of cosmology~\cite{GLASER2017265}.

In group field theory, the emergence of a homogeneous state is tackled by considering condensate states~\cite{Gielen:2016dss}, presented in detail by Steffen Gielen: the excitations, e.g.~above a Fock vacuum, are interpreted as discrete ``atoms of space-time''. The heuristic idea how a smooth, continuous space-time can emerge from this microscopic description is a hydrodynamic one. A large collection of these space-time atoms undergo a phase transition and condense similar to Bose Einstein condensates~\cite{Oriti:2016acw}, such that a macroscopic, effective dynamics emerges from their collective behaviour. In the context of cosmology, one considers a gas of equilateral, uncorrelated building blocks that describe weakly interacting Bose Einstein condensates. In this setup, one can compute expectation values of observables, e.g.~the volume of these building blocks. The dynamics is truncated to the classical equations of motion of the mean field of the condensate, analogous to the Gross-Pitaevskii equation of a Bose Einstein condensate. Remarkably, in this setting the expectation values of observables satisfy effective Friedmann equations~\cite{Oriti:2016ueo}.

Another possible effect of non-perturbative quantum gravity on cosmology are discreteness effects.
In theories where the discreteness is considered as fundamental, such as causal set theory, effects of the discreteness can lead to observable effects and explain certain phenomena.
For example, the randomness inherent in the discrete causal sets can give rise to a cosmological constant of the correct order of magnitude~\cite{Sorkin_2003}.
Since this cosmological constant is no longer constant, it can vary over the age of the universe.
This idea has given rise to phenomenological models that can match the standard model of cosmology and agrees with many of the observables known therein~\cite{zwane_cosmological_2017}.

Another fascinating possibility of cosmological characteristics arising from non-perturbative dynamics was hinted at in the model system of the $2$d orders in causal set theory.
The closest causal set equivalent to the Hartle-Hawking wave function for the early universe is to simulate $2$d orders that are fixed to begin with a single element and to end in a $n$ element anti-chain, the closest causal set equivalent to a spatial hyper surface of fixed volume.
In this model the configurations with the highest likelihood are those that expand rapidly and are very homogeneous~\cite{Glaser:2014dwa}.
While this is a highly simplified model, it shows the possibility to generate features similar to those that have been observed in our universe from non-perturbative dynamics.

\subsubsection{Measuring dimension}

The dimension of space-time is a familiar concept in general relativity. For each point of a $d$-dimensional manifold, we can find a small open region, which we can smoothly map to an open region of $\mathbb{R}^{(1,(d-1))}$ (for Lorentzian signature). As a property of the (topological) manifold, we will refer to this as the topological dimension. While this notion is intuitive in continuum gravity, it is not obvious how to define a dimension in (discrete) quantum gravity. In CDT or spin foam gravity, it is natural to regard the dimension of the fundamental building blocks as the topological dimension. However, whether this ``dimension'' also emerges on large scales is unclear: 4D hypercubes arranged in one long line appear one dimensional on large scales or some building blocks might be degenerate, i.e.~possess vanishing 4-volume. Furthermore, in causal set theory one cannot associate a dimension to discrete space-time events. These difficulties have motivated the definition and investigation of effective dimension measures, that allow us to infer the dimension of space-time, e.g.~via simulations, and potential physical consequences. Indeed, it is an important first test for any approach to quantum gravity, whether these generalized notions of dimension agree with our expectation of four space-time dimensions on large scales.

Moreover, this measured dimension may change with the scale at which space-time is probed, which is further motivation to study such observables.
In general there are several ways to define measures of dimension and all of them have different implications.
One example is the Hausdorff dimension~\cite{2005math......5099S}.
This notion of dimension can be assigned to all metric spaces, via the so-called Hausdorff measure.
It is usually defined for a positive, real parameter $d$ and considers all possible open coverings of the metric space, such that the diameter of each open subset is smaller than $\epsilon$.
The Hausdorff measure with respect to $d$ and $\epsilon$ is then given by the infimum of the sum of all the diameters of the subsets to the power $d$.
To find the Hausdorff dimension, we send $\epsilon \rightarrow 0$ and find the infimum $d$ for which the Hausdorff measure vanishes, which is directly related to how quickly volumes of sets shrink with decreasing diameter.
In quantum gravity, but also random geometries, this notion of dimension is frequently inferred from the exponential growth of volumes with respect to the radius.
Then the Hausdorff dimension is defined as the logarithmic derivative of the volume with respect to the radius, which can change as a function of the radius.

One definition of a scale dependent dimension prevalent in quantum gravity is the spectral dimension.
After first rising to prominence in $4$d simulations of CDT~\cite{Ambjorn_Jurkiewicz_Loll_2005,Benedetti_Henson_2009}, it was also explored in many other theories, e.g. asymptotic safety~\cite{PhysRevD.87.124028},  Ho\v{r}ava-Lifshitz gravity~\cite{Horava_2009}, causal set theory~\cite{ Eichhorn_Mizera_2014,Carlip_2015}, loop quantum gravity~\cite{Modesto:2008jz,Calcagni:2014cza}, spin foams~\cite{Steinhaus_Thurigen_2018} and noncommutative geometry~\cite{Alkofer_Saueressig_Zanusso_2015,BDG_2018}.
This dimension measure is related to studying the heat equation / a diffusion process on space-time.
It crucially depends on the Laplace operator and its spectral properties.
More precisely this dimension measure is defined as the logarithmic derivative of the heat kernel.
For calculations in discrete theories the heat kernel can also be considered as the return probability of a random walker and thus calculated as an average over a sample of random walks.
As a result the spectral dimension encodes how space-time is ordered and thus might reveal interesting consequences for how matter propagates on this geometry, however it is not obvious how to find this connection.
Indeed, Giulia Gubitosi pointed out that the spectral dimension is problematic as a quantity of interest, since it cannot be measured experimentally.
In most approaches it is implemented purely on the geometry, since most computer simulations currently do not include matter.
However all currently conceived experimental measurements of space-time need test particles / test fields.
Hence to define practically observable quantities we will need to work with matter.
As an alternative she suggested the thermal dimension, which tries to define a temperature based on the scaling of thermodynamic properties of matter~\cite{Amelino-Camelia_Brighenti_Gubitosi_Santos_2017}.
This proposal is based on the dimension dependence in the Stefan Boltzmann law, describing the thermal radiation of a theory.
While this is interesting in principle, and they show how it works in Ho\v{r}ava-Lifshitz gravity, where a preferred frame is available, the implementation for a non-perturbative theory is more challenging. Defining a temperature and other thermodynamic quantities in a background independent way can be complicated, a nice discussion of these problems in the context of GFT is given in~\cite{Kotecha:2018gof}.

\subsubsection{Other observables}

In addition to these larger overarching themes that were discussed at lengths and from the perspective of different theories, there were also some interesting observables discussed that are not yet explored in many theories.
One such promising observable is the so-called quantum Ricci curvature~\cite{Klitgaard:2017ebu}; the idea underlying this observable is the following: consider two points in a $d$-dimensional manifold with geodesic distance $\delta$ and imagine each of them to be surrounded by a sphere of radius $\epsilon$.
The points on the sphere are parametrized by a vector from the center to the sphere itself.
Points on the two spheres are related by parallel transporting a vector from one sphere to the other along the geodesic connecting the centers.
The average distance of points on these two spheres depends on the Ricci curvature 2-form (evaluated for the tangent vector of the geodesic connecting the centers), e.g.~if the Ricci curvature is positive the average distance is smaller than $\delta$.
Since this concept is based on parallel transport, it is not straightforwardly applicable to the simplicial geometries underlying (C)DT.
Instead one considers the average distance between all points on the spheres allowing the authors to identify the sign of curvature in constantly curved geometries.
Moreover they have tested it for 2D-(E)DT with spherical topology revealing a positively curved geometry modeled as a 5D sphere emphasizing the highly non-classical and fractal geometries in this model~\cite{Klitgaard:2018snm}.
It will be interesting to see the behaviour of this observable in 4D CDT and whether it can be translated to other approaches of quantum gravity.

Another interesting route to explore is holography, more precisely the deep relation between a theory in the bulk and the theory on the boundary.
This is most prominently represented in the continuum by the infamous AdS/CFT correspondence~\cite{Maldacena:1997re}.
Naturally it is an interesting question to ask whether these ideas can be generalized to non-perturbative approaches to quantum gravity and what the corresponding boundary theories might be.
A very interesting calculation has been performed for the Ponzano-Regge model of 3D spin foams, studying the partition function and dual boundary theory of the twisted solid torus~\cite{Dittrich:2017hnl,Dittrich:2017rvb} (see also a similar calculation for linearized Regge calculus~\cite{Bonzom:2015ans}).
Strikingly the results are consistent with results from perturbative quantum field theory in the continuum~\cite{Barnich:2015mui,Oblak:2015sea} and the characters of the BMS group are recovered.
In addition there have been several derivations for holographic entanglement entropy, more precisely the Ryu-Takayanagi formula for Renyi entropy, where the entropy (of the boundary theory) associated to a boundary subsystem is proportional to the minimal bulk area attached to this section of the boundary~\cite{Han:2016xmb,Chirco:2017vhs}.

\subsection{Numerical methods in Quantum Gravity}\label{subsec:num}

Using physical intuition to develop our algorithms can lead to massive improvements in speed.
At our workshop we were introduced to two algorithms employing this, the chimera algorithm for numerical loop quantum cosmology and the transfer matrix algorithm for CDT.

\subsubsection{The chimera algorithm}
Parampreet Singh told us about the chimera algorithm developed in loop quantum cosmology. One of the key features of loop quantum cosmology is the resolution of the Big Bang singularity at the origin of the universe via a Big Bounce~\cite{PhysRevLett.96.141301}. The vital dynamics responsible for this result is encoded in the quantum Hamiltonian constraint, which is a difference equation with uniform discretisation in volume. Indeed, for small volumes and large space-time curvature this dynamics significantly deviates from the classical dynamics given by a Wheeler-DeWitt differential equation. However, for large space-time volume and small curvature the quantum and classical dynamics agree very well. This is the fundamental idea underlying the chimera algorithm~\cite{Diener_Gupt_Singh_2014}.

Difference equations, which describe the evolution in the deep quantum regime, are much more costly to compute compared to ordinary differential equations. This issue is emphasized as soon as the quantum states, whose evolution is studied, are not sharply peaked on classical configurations. Thus the chimera algorithm introduces a hybrid lattice, where quantum evolution is only performed at small volumes and classical dynamics takes over for large volumes. The intermediate region is carefully chosen for the results to match. That way the numerical costs are drastically reduced and can be spent instead on studying the evolution of more general quantum states~\cite{Diener_Gupt_Singh_2014}. It will be interesting to explore whether this idea of a hybrid algorithm can be adapted in other approaches as well, e.g.~spin foam models.

\subsubsection{Transfer matrix approach}
The transfer matrix approach is well known as a method to solve e.g. the Ising model in $2$d. It is an analytic method based on splitting a problem into layers, e.g. time slices, calculating the dynamics of a single layer and then combining consecutive layers by convolution.
This method also works to analytically solve CDT in $2$ dimensions.
In higher dimensions CDT can only be explored using computer simulations, however the foliated structure still makes it a prime candidate for the transfer matrix approach.
Andrzej G\"{o}rlich explained how this insight and a clever numerical implementation of the transfer matrix approach were used in~\cite{transfer_2014}.
In his algorithm he measures the transfer matrix between slices of fixed size, such that he only has to simulate two slices of geometry, instead of the entire universe.
This allows for more focused measurements, in particular improving the precision in measuring off-diagonal elements of the transition amplitude immensely.
The transfer matrix approach also had another, unexpected, payoff in showing that what was before considered a single de~Sitter like phase, called phase C, splits into two different phases: one that is de~Sitter like, and one with alternating large and small spatial slices, called the bifurcation phase~\cite{transfer_2014}.

\subsubsection{Markov Chain Monte Carlo simulations}
Other improvements in code are less about understanding the physical situation of the problem, and more about understanding the idiosyncrasies of a particular simulation.
The most used tool to calculate a path integral using computer simulations are Markov Chain Monte Carlo (MCMC) simulations.
How this algorithm is applied in their respective approaches to quantum gravity was explained by  Andrzej G\"orlich, Jack Laiho, and Sumati Surya.
In these the ensemble of geometries is sampled with a frequency proportional to the weight of the configurations in the path integral, which makes it easy to calculate averages of observables directly from the simulations.
A Markov chain, is a chain in which the likelihood to transition between two states only depends on these two states. One algorithm to generate such a chain is the Metropolis-Hastings algorithm.
This algorithm generates a Markov Chain by proposing a new state as a function of the old one.
The function proposing these states depends on the theory used, e.g. in dynamical triangulations it is given by Pachner moves, which locally change the triangulation~\cite{PACHNER1991129}.
The probability to accept a proposed move then depends on the weight of the geometry in the path integral, given by $e^{-S}$ with $S$ the action of the theory.
One important feature is that a new state will always be accepted if it has a higher weight, but even states with a lower weight can still be accepted with a probability proportional to $e^{S_{old}-S_{new}}$.
This makes it possible to prove that, if the moves are ergodic, the Metropolis-Hastings algorithm will find a global minimum of the action, if run sufficiently long.
Unfortunately the convergence towards this can be very slow, particularly close to phase transitions, since most proposed moves will have a very low probability of being accepted.
This is known as critical slowing down and is related to the divergence of the correlation length arising there.

\subsubsection{Parallel rejection}
One algorithm to overcome critical slowing down is the parallel rejection algorithm, discussed in Andrzej G\"orlich's talk.
In general, MCMC simulations are difficult to parallelize, particular in gravity systems, since changes in the value of the action are non-local, hence proposed moves are not independent and need to be calculated sequentially.
In practice, this means that most simulations are ``naively parallelized'' by just starting the simulations for several different points in the phase diagram, different parameter values, at the same time on different cores.
Parallel rejection is an algorithm that does actual parallelization, for at least some regions of the phase diagram, where it can substantially speed up the algorithm.
In regions of the parameter space where the acceptance rate of moves is particularly low,  parallel rejection proposes and calculates multiple moves at the same time, on different cores.
Once one of them is accepted (which can be $\sim 1 \%$ or less of proposed moves) the geometry is updated and the parallel rejection restarted.
This can drastically reduce the time in which the code remains in a given configuration in these regions of the phase diagram~\cite{Laiho:2016nlp}.

\subsubsection{Adopting methods from other fields}
Another interesting option is  to start using tools from other areas of science, in particular from computer science.
There are many techniques that are solidly established in other fields but have not been widely adapted in numerical quantum gravity yet.
For example, in QCD the default algorithm for simulations is not the Metropolis Hastings algorithm, instead the algorithms in use are hybrid Monte Carlo~\cite{Neal_2012} explained in Jack Laiho's talk.
In these the step of proposing a new configuration is guided by a supplementary Hamiltonian function. This Hamiltonian function is defined with respect to the probability distribution we wish to sample from and introduces fictitious momenta. While the momenta are randomly updated, a step in the configuration variables is chosen via a Metropolis algorithm with respect to the Hamiltonian equations of motion, which results in a faster convergence of results.
This ``Hamiltonian'' is not be confused with an energy functional or the Hamiltonian constraint in gravity and serves the purpose to optimize the updating of configurations.
The drawback of this method is that it requires continuous configurations, which makes it unsuitable for many proposals in quantum gravity.

Parallel tempering, discussed by Andrzej G\"orlich, also known as replica exchange MCMC sampling, is very useful when the configurations generated, e.g.~from a Metropolis algorithm, are highly auto-correlated, that is correlated with previously generated configurations~\cite{Newman_Barkema_1999}. Such correlated systems may suffer from critical slowing down, in which the system is unlikely to leave said configuration via the proposed updates. To avoid this, the principle idea of parallel tempering is to start several processes with different model parameters and exchange the configurations at some point. That way regions in configuration space that are rarely explored for certain parameters become accessible, improving the accuracy of the simulation. Often it is proposed that the parameters only slightly vary. The probability to exchange the configurations has to satisfy the detailed balance condition. Crucially, this algorithm significantly reduces the auto-correlation time, i.e.~the time it takes for configurations of the same Markov chain to become statistically independent.

In recent years, deep learning has emerged as a powerful method to analyze and search for patterns in large amounts data. Image recognition is a particularly impressive example. Naturally we would like to apply these methods to quantum gravity, e.g.~to examine data generated in Monte Carlo simulations. In a nutshell, the idea of deep learning is to find an optimal function that quickly returns a desired output from a given large input. Deep neural networks are usually modelled to have an in- and output layer, chosen according to data and desired output. Between these layers, one implements several hidden layers, where each neuron in a hidden layer is connected to all neurons in the previous and following layer. These connections simply encode linear algebra operations on the data. Then some of the data is used for training, i.e.~these linear algebra operations get optimized to minimize a cost function. In supervised learning, where we know the desired result for a given sample, we would optimize the neural network to reproduce the already known answer.
This is a particularly powerful approach when it comes to classification problems, e.g. recognizing handwriting or in quantum gravity it might help us to sort geometries with different properties.
This approach bears an enormous potential, yet comes with some obvious drawbacks. Indeed, it is not obvious how to design a deep neural network that can successfully analyze a given data set. Moreover, even once we have successfully trained a neural network, it might not be obvious what the computer has learned, limiting our interpretation and understanding of the problem.
Another problem is that, at least for the easiest to apply algorithms with the clearest outcomes, called supervised learning, we need to label the data set beforehand.
This was beautifully demonstrated by Will Cunningham in his talk: he uses causal sets of known dimension, either $d=2$, $3$ or $4$, to train a neural network to determine the dimension of the causal set.
This is an interesting toy model, that demonstrates the opportunity and the challenge of machine learning at the same time.
The characterization of the causal sets he obtained through the algorithm could have equally well been done using many tools that have been developed in causal set theory, e.g. the Myrrheim-Meyer dimension~\cite{Meyer_1988} or the interval abundance~\cite{Glaser_Surya_2013}, which are fast and simple to use.
On the other hand, these tools took time to develop and relied on our deep understanding of the problem, while the computer was obviously not aware of these and still able to solve it.

In general, quantum gravity, in particular in approaches that heavily use Monte Carlo simulations, offers many opportunities to apply machine learning.
It will not always be possible to label the large data sets generated by Monte Carlo simulations and unravel all of their ``hidden'' information. Hence, using machine learning to search for structure within, and to possibly identify new observables, is a worthwhile endeavour.

\section{Roadmap}\label{sec:roadmap}
\begin{figure}
    \centering
    \includegraphics[width=\textwidth]{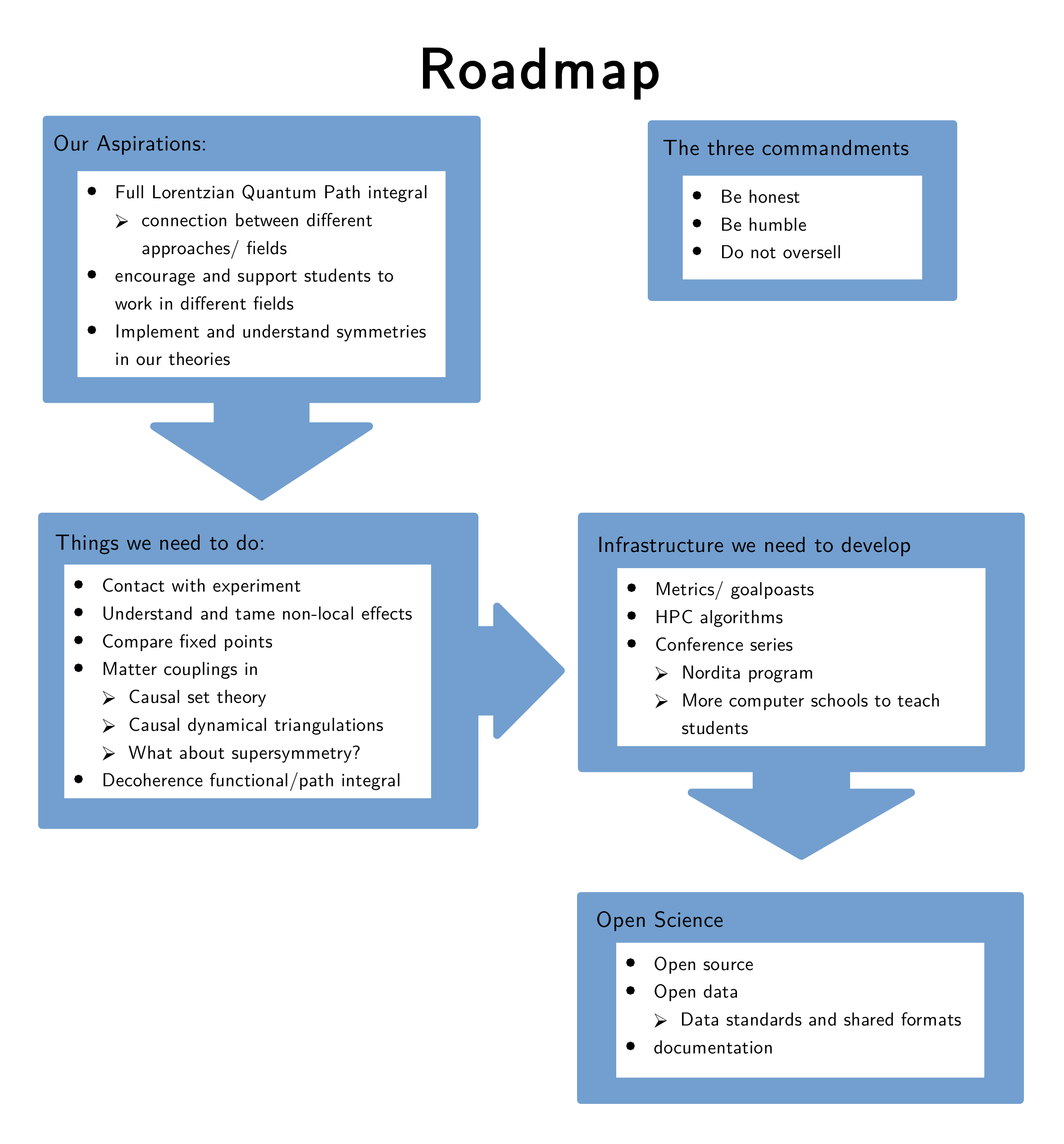}
\caption{A slightly cleaned up version of the flowchart we created in the concluding discussion.}
    \label{fig:flowchart}
\end{figure}
We ended our workshop with a roadmap discussion, in which we began charting the future course of numerical quantum gravity.
One outcome of this discussion is a flowchart, summarizing the discussions of the workshop and pointing towards questions for future consideration reproduced in Figure~\ref{fig:flowchart}.

\subsection{Open science}
One point of discussion which received particular emphasis in generating the roadmap was the desirability of conducting open science.
In the context of computational quantum gravity this would boil down to two points; open source code and open data.

Developing, optimizing and running code is an integral part of numerical research. In quantum gravity, currently most code is in principle available, but requires interested researchers to reach out and ask the authors for access.
While this allows the authors to somewhat keep control of who has a current version of their code, it would be desirable for the development of the field to make their code open source.
Open source means that the code is publicly available: anyone who is willing to improve the code or use and adapt it for their own research can do so without seeking permission of the authors.

There exist many good solutions for storing and distributing open source code, for example the platform github\footnote{\url{https://github.com/}}.
This website is built around the version control software git\footnote{\url{https://git-scm.com/}}.
Git keeps track of any changes made to the code, and shows a history of the repository with all the changes made in various commits\footnote{In a commit, the author submits the changes made to the code to the repository.}. This makes it possible to revert to previous versions and makes it straightforward to work with multiple people on the same project.
Using git, anyone can download and use the code by cloning the repository. Then they can also commit changes to the code, which must be approved by the owners of the repository.

We believe that this practice, which is standard even in closed source software development, has many advantages and its adoption by the quantum gravity community would boost the development of our numerical efforts.
Indeed, authors deserve credit for their code, where the ideas and work that went into developing it are often not reflected in papers. Hence, an open source strategy makes this readily accessible, which makes it more easy for other researchers to contribute to the field and adopt ideas. Moreover, it makes research more credible and reliable, since the tools are readily available, to verify results. As a last point, open source is a good motivation to document and explain one's code, such that is usable for other people. That way, even once a researcher has left the field, their code is still available.

While there are several platforms and tools available to share and publish code, it is much more difficult to publish or exchange large amounts of data. Indeed, having public access to data generated in computer simulations is desirable for many reasons. Being able to recreate and confirm results greatly enhances the credibility of one's research. Moreover, it allows other researcher, e.g.~from a different field like phenomenology, to analyze the data and use it for their own research. As a final point, large scale numerical simulations are costly and not every interested researcher has access to advanced numerical resources, e.g.~in developing countries. Openly available data sets allow more people to learn and contribute to the field, e.g.~students.

We envision different types of data to be uploaded, depending on the approach to quantum gravity. For approaches such as CDT and causal sets that rely on MCMC simulations, one option would be to upload the samples generated in the simulations allowing other researchers to investigate them for new patterns or calculate observables. In spin foam gravity uploading exact values of spin foam amplitudes, which can then be readily used in other calculations, would be another straightforward example. All of the uploaded data should be reusable by other researchers and be accompanied by a documentation on how to use the data and / or contain a short program to demonstrate how to read in the data. Also modern efficient file formats should be used, like HDF5 or CSV, in particular for large files. On that note, uploaded files should be compressed to reduce internet traffic, and if the amount of data is particularly large, it should be split into smaller files.

Thus at the end of the workshop, a plan was hatched to implement a quantum gravity open data repository. Together with other participants of the workshop, Benjamin Bahr, William Cunningham and Bianca Dittrich, as well as Erik Schnetter and Dustin Lang, we are actively developing the concept and realization of an open data initiative for the field of quantum gravity. The current plan is for this repository to be open to all numerical approaches, with a wiki-style website that allows authors to easily add data and link it to their papers on arXiv.
Moreover, a DOI should be automatically assigned to each published dataset to make it straightforwardly citable.
We are currently in the process of discussing the exact format and procuring funding for this endeavour\footnote{Any recommendations for sources of funding, or inspirations for how to set up such a project are very welcome}, the working title is ``The encyclopedia of quantum geometries''.

\subsection{Future}
To keep the discussion alive we plan to apply for funds and organize follow-up workshops and schools, currently the most likely schedule will be to have a school one year\footnote{For schools we would imagine a format similar to that of \href{https://www.perimeterinstitute.ca/video-library/collection/making-quantum-gravity-computable}{``Making quantum gravity computable''} at the Perimeter institute.} and then hold workshops in alternating years.
All participants seemed excited by the prospect of future such workshops, and there is a large number of interested parties that could not make it this year but have asked to be notified about future plans.
One particularly exciting possibility would be to organize a \href{https://www.nordita.org/events/programs/index.php}{Nordita programme} on quantum gravity embedding the workshop and the school into one longer event.

We hope to have done justice to all participants and their many brilliant contributions to this conference and hope there will be many further conferences on this exciting subject.
So, until then be honest, be humble and do not oversell.

\begin{acknowledgements}
Traditionally conferences and workshops disseminate their results through proceedings.
This has the disadvantage of requiring the organizers to chase down the speakers, while it might still miss important points raised in discussions.
Hence we decided to write a recap of it ourselves, and would like to thank all participants for making the conference a huge success and giving us so much to think and write about.
We would also like to thank Nordita for hosting the workshop and supporting us in the organization.
L.G. is supported by the People Programme (Marie Curie Actions) H2020 REA grant agreement n.706349 ”Renormalisation Group methods for discrete Quantum Gravity”, which also supported speaker travel for the conference.
This research was supported in part by Perimeter Institute for Theoretical Physics. Research at Perimeter Institute is supported by the Government of Canada through Innovation, Science and Economic Development Canada and by the Province of Ontario through the Ministry of Research, Innovation and Science.
The conference was also supported by COST Action MP1405 QSpace - `Quantum Structure of Spacetime'.
\end{acknowledgements}
\bibliography{bibliography}
\end{document}